\documentclass[letter]{aa} 

\usepackage{graphicx}
\usepackage{txfonts}
\usepackage{hyperref}
\begin{document}

   \title{A dynamical mass for GJ 463 b: A massive super-Jupiter companion beyond the snow line of a nearby M dwarf
   }

   \author{A. Sozzetti
          \inst{1}
          }

    \institute{INAF - Osservatorio Astrofisico di Torino, Via Osservatorio 20, I-10025 Pino Torinese, Italy\\
              \email{alessandro.sozzetti@inaf.it}
             }

   \date{Received XXX; accepted XXX+1 day}

  \abstract{We determined the full orbital architecture and true mass of the recently Doppler-detected long-period giant planet GJ 463 b using the HIPPARCOS-Gaia proper motion anomaly in combination with the available radial velocities, constraints from the knowledge of the spectroscopic orbital parameters, and supplementary information from a sensitivity analysis of Gaia Data Release 3 astrometry. We determined an orbital inclination $i_b=152^{+2}_{-3}$ deg (for a prograde orbit) and a mass ratio $q=0.0070\pm0.0007$, corresponding to a true mass of the companion $M_b=3.6\pm0.4$ M$_\mathrm{Jup}$. True mass determinations for a super-Jupiter companion at intermediate orbital separations beyond the snow line around low-mass stars ($M_\star\leq 0.5$ M$_\odot$) are a rare occurrence. Its existence is possibly explained in the context of disk-instability models of planet formation.}

\authorrunning{A. Sozzetti}
\titlerunning{The true mass of GJ 463 b}
   \maketitle
%

\section{Introduction} \label{sec:intro}

The frequency of close-in ($a\lesssim 1$ au) gas giant planets around low-mass M dwarfs ($M_\star\lesssim 0.7$ $M_\odot$) is known to be significantly lower than that of Jupiter- and super-Jupiter-mass companions to solar-type stars (e.g. \citealt{Endl2006,Bonfils2013,Gan2023}). At wider separations, the convergent view from Doppler, microlensing, and direct imaging surveys is that gas giants appear more common (e.g. \citealt{JOhnson2010,Gould2010,Cassan2012,Montet2014}), similar to the trend observed for solar-type hosts (e.g. \citealt{Wittenmyer2020}, and references therein). However, occurrence rate estimates still carry large uncertainties, and it remains unclear whether the frequency of cold Jupiters around M dwarfs is lower than that of the same population orbiting earlier-type stars (e.g. \citealt{Pinamonti2022}, and references therein). More detections of long-period giant planets orbiting M dwarfs are therefore highly desirable in order to improve studies on the demographics of this component of the planetary population. The nearby ($d=18.4$ pc) early-M dwarf GJ463 (Ross 690, HIP 60398) was recently identified, through long-term radial-velocity (RV) monitoring, to host a planetary companion on a $\sim10$ yr orbit with a minimum mass of $M_b\sin i \sim1.6$ M$_\mathrm{Jup}$ \citep{Endl2022}. In the discovery paper, \citet{Endl2022} qualitatively discuss the constraints on the true mass of GJ 463 b based on evidence from HIPPARCOS-Gaia absolute astrometry and a Gaia Data Release 3 (DR3, \citealt{Vallenari2022}) diagnostic of the departure from a good single-star fit to  Gaia-only astrometry -- the re-normalised unit weight error (RUWE). For this study, we effectively used the astrometric acceleration -- hereafter the proper motion anomaly (PMA) -- measured by \citet{Kervella2022} in combination with the available RVs and constraints on the orbital parameters from the RV solution presented in \citet{Endl2022} to determine actual values for the inclination of the orbital plane $i_b$, the longitude of the ascending node $\Omega_b$, and the mass ratio $q$. In combination with an analysis of the Gaia DR3-level sensitivity to orbiting companions, this allows us to provide a direct  measurement of the true mass of GJ 463 b. 

\section{Analysis}

\subsection{Spectroscopy and absolute astrometry}

Table \ref{tab:parameters} summarises all the parameters and data used in our analysis. The spectroscopic orbital elements and primary mass have been taken from \citet{Endl2022}. The Gaia DR3 RUWE, $G$ mag, parallax $\varpi$, and colour $BP-RP$ values have been taken from \citet{Vallenari2022}. The PMA vector components ($\Delta\mu_\alpha$ and $\Delta\mu_\delta$) at the mean epochs of the HIPPARCOS and Gaia DR3 catalogues have been taken from \citet{Kervella2022}. These quantities were obtained subtracting from the quasi-instantaneous proper motions of the two catalogues' the long-term proper motion vector defined as the ratio of the positional difference between the two catalogues to the time baseline ($\sim25$ yr). As the latter quantity is a factor $\sim2.5$ larger than the orbital period of GJ 463 b, it can be considered as a good representation of the tangential velocity of the barycentre of the system. The observed $\Delta\mu$ values reported in Table \ref{tab:parameters} therefore are expected to contain only information on the orbital motion of GJ 463 b. 

\begin{table*}
    \centering
        \caption{Parameters of the GJ 463 planetary systems used in the analysis. The publicly available RVs from \citet{Endl2022} are not shown. \label{tab:parameters}
}
        \begin{tabular}{lcc}
    \hline
    \hline
    \noalign{\smallskip}
    Parameter     &  Value &  Reference\\
    \noalign{\smallskip}
    \hline
    \noalign{\smallskip}
    \noalign{\smallskip}
    \noalign{\smallskip}
    Stellar mass $M_\star$ ($M_\odot$) & $0.49\pm0.02$ & \citet{Endl2022}  \\
    \noalign{\smallskip}
    Parallax $\varpi$ (mas)  & $54.447\pm0.019$ & \citet{Vallenari2022}  \\
    \noalign{\smallskip}
    Gaia DR3 RUWE & $1.407$  & \citet{Vallenari2022} \\
        \noalign{\smallskip}
    $G$ mag & $10.552$  & \citet{Vallenari2022} \\
        \noalign{\smallskip}
    Gaia colour $BP-RP$ & $2.396$  & \citet{Vallenari2022} \\
        \noalign{\smallskip}
    RV semi-amplitude $K$ (m s$^{-1}$) & $33.3\pm3$  & \citet{Endl2022} \\
        \noalign{\smallskip}
    Orbital period $P$ (d) & $3448^{+110}_{-88}$  & \citet{Endl2022} \\
        \noalign{\smallskip}
    Epoch of periastron $T_0$ (BJD) & $2454457^{+82}_{-90}$  & \citet{Endl2022} \\
        \noalign{\smallskip}
    Eccentricity $e$ & $0.09^{+0.18}_{-0.05}$  & \citet{Endl2022} \\
        \noalign{\smallskip}
    Argument of periastron $\omega$ (rad) & $-1.2^{+0.7}_{-0.3}$  & \citet{Endl2022} \\
        \noalign{\smallskip}
    Semi-major axis $a_b$ (au) & $3.53\pm0.07 $  & \citet{Endl2022} \\
        \noalign{\smallskip}
    Minimum companion mass $M_b\sin i$\, (M$_\mathrm{Jup})$ & $1.55\pm0.15$  & \citet{Endl2022} \\
        \noalign{\smallskip}
    HIPPARCOS (epoch 1991.25) $\Delta\mu_\alpha$ (mas yr$^{-1}$) & $+4.330\pm2.001$  & \citet{Kervella2022}\\
        \noalign{\smallskip}
    HIPPARCOS (epoch 1991.25) $\Delta\mu_\delta$ (mas yr$^{-1}$) & $-0.703\pm2.141$  & \citet{Kervella2022} \\
        \noalign{\smallskip}
    Gaia (epoch 2016.0) $\Delta\mu_\alpha$ (mas yr$^{-1}$) & $-0.641\pm0.075$  & \citet{Kervella2022} \\
        \noalign{\smallskip}
    Gaia (epoch 2016.0) $\Delta\mu_\delta$ (mas yr$^{-1}$) & $+0.311\pm0.058$  & \citet{Kervella2022} \\
        \noalign{\smallskip}
    \hline
    \end{tabular}
\end{table*}

\subsection{Constraints on $i$, $\Omega$, and GJ 463 b's true mass}

The analysis builds and expands upon the methodology described in \citet{Damasso2020}. We fitted the PMA astrometric data together with the publicly available RVs of GJ 463, adopting the spectroscopically determined orbital parameters as constraints. In the combined RV+PMA model, we explored the possible values of the two orbital parameters of GJ 463 b that can be constrained by the PMA astrometric data ($i_b$ and $\Omega_b$) and the mass ratio $q$.

A differential evolution Markov chain Monte Carlo (DE-MCMC) algorithm \citep{TerBraak2006,Eastman2013} was utilised, with uniform priors on $\cos(i_b)$ and $\Omega_b$ over the allowed ranges for both prograde and retrograde motion, as well as a broad uniform prior on $q$ (see Table \ref{tab:bestfit}). Uninformative priors were utilised for the remainder of the model parameters. 
The PMA model was built by taking averages over the actual observing windows of HIPPARCOS and Gaia DR3. To this end, we utilised the exact times of HIPPARCOS observations of GJ 463 available in the HIPPARCOS-2 Epoch Photometry Annex \citep{vanLeeuwen2007}, and obtained a close representation of the actual Gaia transit times from the Gaia Observation Forecast Tool (GOST)\footnote{\url{https://gaia.esac.esa.int/gost/index.jsp}}. Observing window averaging is necessary to cope with the 'smearing' effect of the orbital motion due to the fact that the proper motions are averaged over the data-taking intervals of both HIPPARCOS and Gaia, and a loss in sensitivity is non-negligible even in the case of an orbital period comparable to that of GJ 463 b (see Fig. 2 of \citealt{Kervella2019}). The final adopted likelihood is $\ln \mathcal{L} = -0.5\,\left(\chi^2_\mathrm{RV}+\chi^2_\mathrm{\Delta\mu_\alpha} + \chi^2_\mathrm{\Delta\mu_\delta}\right)$. 

\begin{figure}
\centering
\includegraphics[width=0.95\columnwidth]{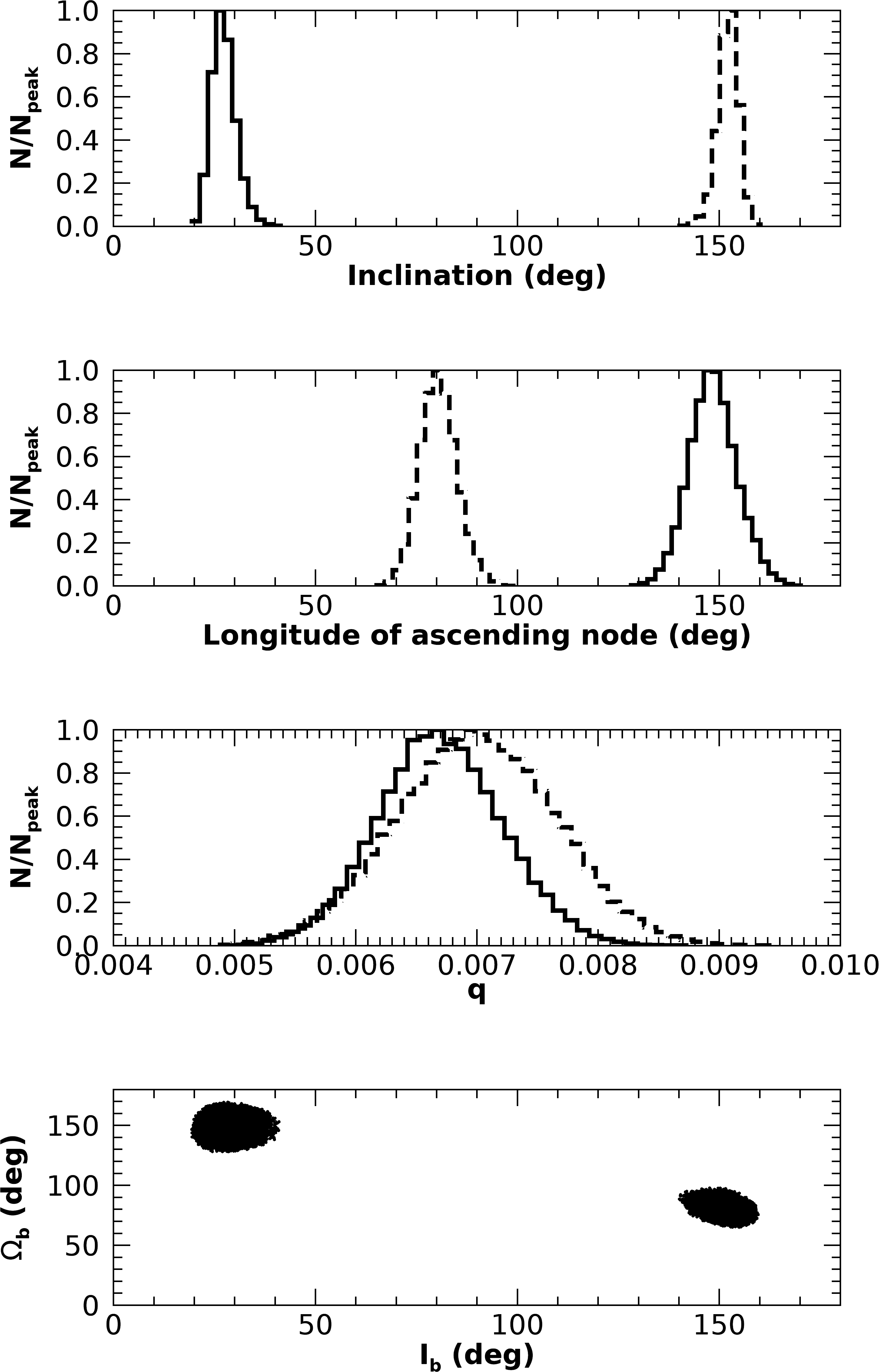} 
 \caption{Constraints on inclination, longitude of the ascending node, and mass ratio from the combined RV+PMA orbital fit. Top and central panels: Posterior distributions for $i_b$, $\Omega_b$, and $q$ for the direct (dashed histograms) and retrograde solutions (solid histograms). Bottom panel: Joint posterior distributions for $i_b$ and $\Omega_b$.}
\label{fig:gj463_pma}
\end{figure}

In the DE-MCMC analysis, after removal of 10\% of the steps corresponding to the burn-in phase (e.g. \citealt{TerBraak2006}), the medians of the posterior distributions of $i_b$, $\Omega_b$, and $q$ were adopted as the central values of the parameters, while the $1\sigma$ uncertainties on the model parameters were obtained evaluating the $\pm34.13$ per cent intervals of the posteriors. 
The best-fit values of $i_b$, $\Omega_b$, and $q$ for both a prograde and a retrograde orbit are reported in Table \ref{tab:bestfit}, while we show their individual posterior distributions and the corresponding joint posteriors between $i_b$ and $\Omega_b$ in the four panels of Fig. \ref{fig:gj463_pma}. The data at hand did not allow to resolve the ambiguity between prograde and retrograde motion, but the outcome of the analysis clearly points towards the identification of an inclination not far from face-on for GJ 463 b. In the prograde and retrograde solutions, we found $i_b=152^{+2}_{-3}$ deg and $i_b=27^{+3}_{-3}$ deg, respectively. For the prograde solution, the mass ratio is $q=0.0070\pm0.0007$, with a corresponding derived value of true mass for GJ 463 b $M_b=3.6\pm0.4$ M$_\mathrm{Jup}$. The central value of $M_b$ identifies the companion orbiting GJ 463 as a massive super-Jupiter. As a cross-check, we performed the same analysis as above using the values of the PMA vectors from the HIPPARCOS-Gaia catalogue of accelerations constructed by \citet{Brandt2021}, obtaining virtually identical results. 

We corroborated our findings by performing a sensitivity analysis of Gaia DR3 astrometry to companions of a given mass and orbital period based on the RUWE statistic. Its value, 1.407, is exactly at the threshold above which a single-star model fails to satisfactorily describe the data (e.g. \citealt{Lindegren2018,Lindegren2021}). We then followed an approach similar to \citet{Belokurov2020} and \citet{Penoyre2020} to investigate the range of orbital separations and companion masses that would induce excess astrometric
residuals with respect to a single-star model, therefore producing RUWE values larger than the one reported. Using the values of the observing times, along-scan parallax factors, and scan angles obtained from the GOST tool and the nominal Gaia DR3 astrometric parameters, we created synthetic Gaia along-scan observations. We added, linearly, the effects of orbital motion produced by companions with masses in the range $1-40$ M$_\mathrm{Jup}$ and a semi-major axis in the range $1-10$ au (using the parallax and primary mass values from Table \ref{tab:parameters}). For each $M-a$ pair, we generated one hundred random realisations of the other orbital elements, all drawn from uniform distributions encompassing their allowed intervals. Finally, we added Gaussian measurement uncertainties appropriate for a $G=10.5$ mag source (following \citealt{Holl2022}). A single-star model was fitted to the data, and for each mass-separation pair we recorded the fraction of systems with RUWE $>1.407$. 

\begin{table}[ht!]
    \centering
        \caption{Fitted and derived parameters for GJ 463 b from the combined RV+astrometry analysis. \label{tab:bestfit}
}
        \begin{tabular}{lcc}
    \hline
    \hline
    \noalign{\smallskip}
    Parameter     &  Prior &  Value\\
    \noalign{\smallskip}
    \hline
    \noalign{\smallskip}
    \noalign{\smallskip}
    &\textit{Prograde solution} &\\
    \noalign{\smallskip}
    \noalign{\smallskip}
    \hline
    \noalign{\smallskip}
    \noalign{\smallskip}
    $i_b$ [deg] & $\cos(i_b),\,\mathcal{U}(0.0,180.0)$ &  $152^{+2}_{-3}$ \\
    \noalign{\smallskip}
    $\Omega_b$ [deg] & $\mathcal{U}(0.0,360.0)$  & $80^{+6}_{-5}$  \\
    \noalign{\smallskip}
    $q$  & $\mathcal{U}(0.0,0.1)$  & $0.0070\pm0.0007$  \\
    \noalign{\smallskip}
    $M_b$ [M$_\mathrm{Jup}$] & (derived) & $3.6\pm0.4$  \\
    \noalign{\smallskip}
    \noalign{\smallskip}
    \hline
    \noalign{\smallskip}
    \noalign{\smallskip}
    &\textit{Retrograde solution} &\\
    \noalign{\smallskip}
    \noalign{\smallskip}
    \hline
    \noalign{\smallskip}
    \noalign{\smallskip}
    $i_b$ [deg] & $\cos(i_b),\,\mathcal{U}(0.0,180.0)$ & $27^{+3}_{-3}$  \\
    \noalign{\smallskip}
    $\Omega_b$ [deg] & $\mathcal{U}(0.0,360.0)$ & $148^{+6}_{-5}$  \\
    \noalign{\smallskip}
    $q$  & $\mathcal{U}(0.0,0.1)$  & $0.0067\pm0.0005$  \\
    \noalign{\smallskip}
    $M_b$ [M$_\mathrm{Jup}$] & (derived) &  $3.4\pm0.3$ \\
    \noalign{\smallskip}
        \noalign{\smallskip}
    \hline
    \end{tabular}
\end{table}

   \begin{figure}
   \centering
   \includegraphics[width=0.95\columnwidth]{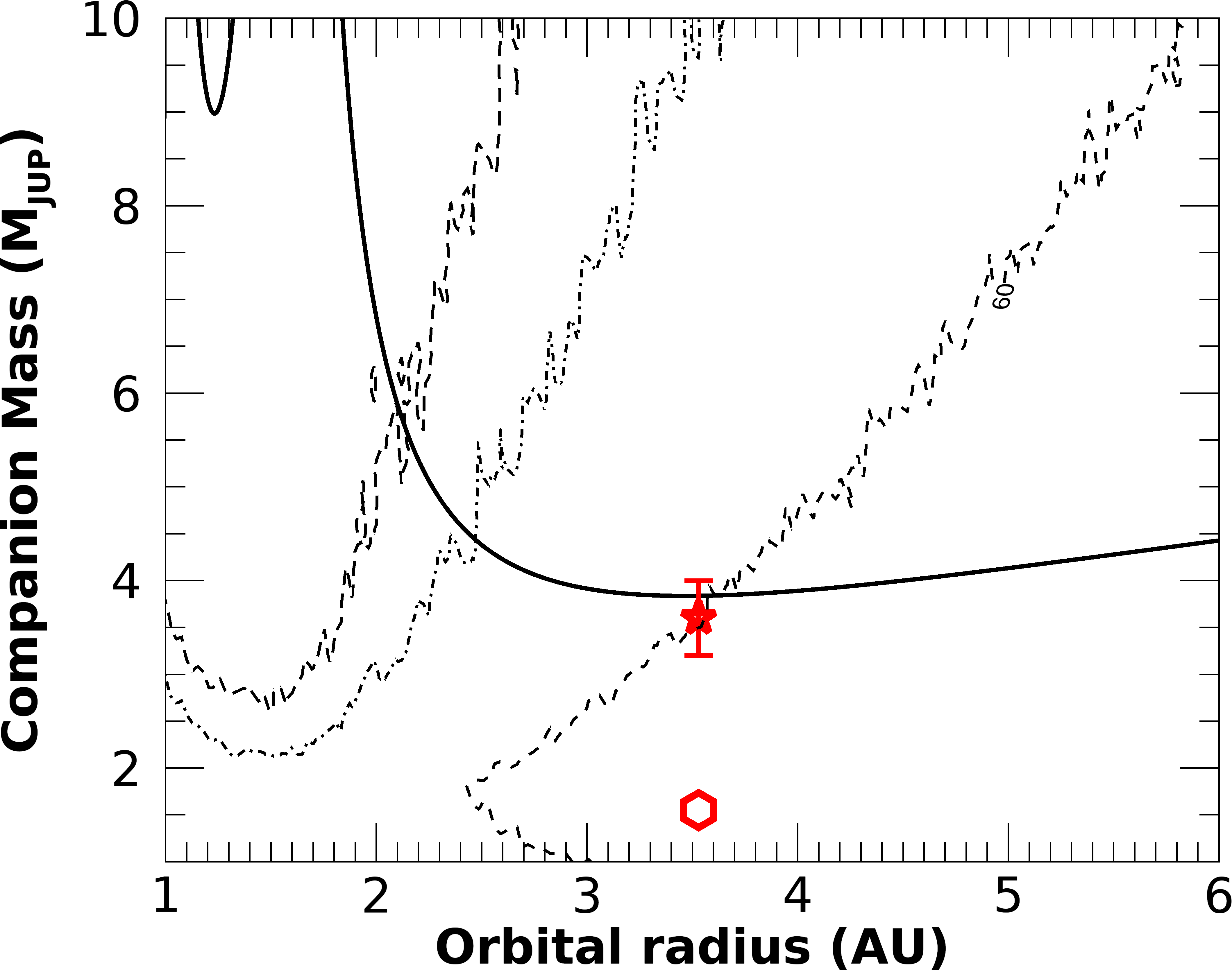}
      \caption{HIPPARCOS-Gaia PMA and Gaia DR3 sensitivity to companions of a given mass (in M$_\mathrm{Jup}$) as a function of the orbital semi-major axis (in au) orbiting GJ 463. Dashed, dashed-dotted, and long-dashed lines correspond to iso-probability curves for 60\%, 95\%, and 99\% probability of a companion of given properties to produce RUWE $> 1.407$. The solid line identifies the combinations of mass and separation explaining the observed $\Delta\mu$ PMA at the mean epoch of Gaia DR3. The red hexagon and star indicate the $M_b\sin i$ value from \citet{Endl2022} and the $M_b$ value for the prograde solution obtained in this work, respectively. 
              }
         \label{fig:sensitivity}
   \end{figure}

Fig. \ref{fig:sensitivity} shows iso-probability contours in $M-a$ space. For example, above the highest curve, there is $>99\%$ probability that a companion of a given $M$ and $a$ would induce a RUWE value larger than the one reported. Fig. \ref{fig:sensitivity} also shows the PMA sensitivity curve based on Eq. 15 of \citet{Kervella2019}. Finally, we report the \citet{Endl2022} minimum-mass value for GJ 463 b and the best-fit true-mass value obtained in our analysis. The HIPPARCOS-Gaia PMA sensitivity curve indicates that, at the orbital separation of GJ 463 b, a companion inducing a statistically significant PMA should have a mass of the order of approximately 4 M$_\mathrm{Jup}$. This is indeed what we have found in our DE-MCMC analysis. The Gaia DR3 sensitivity curve shows that a companion with $a\sim3.5$ au needs to have a mass at least of the order of $8-10$ M$_\mathrm{Jup}$ in order to have a high probability ($\gtrsim90\%$) of being the one responsible for the observed RUWE. There is still a $\sim60\%$ probability that the RUWE value stems from the unmodelled orbital signal due to GJ 463 b, but one could wonder whether the full extent of the measured excess scatter in the post-single-star fit to Gaia DR3 astrometry for GJ 463 can indeed be interpreted solely in terms of the presence of the companion discussed here.

\begin{figure}
   \centering
   \includegraphics[width=0.95\columnwidth]{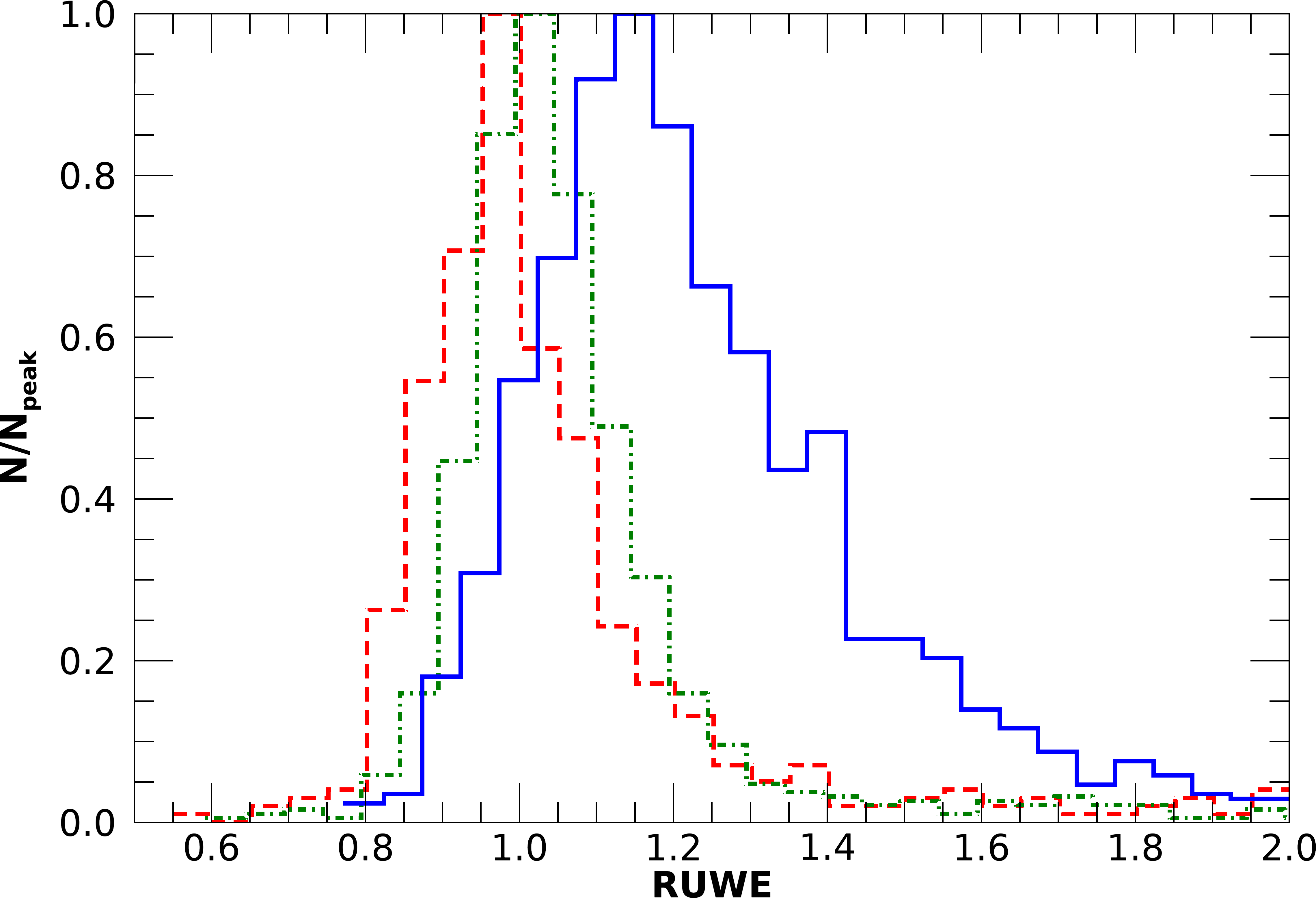}
      \caption{Gaia DR3 RUWE distribution for different stellar sub-samples. Solid blue histogram: Sources with $G<12$, $2<BP-RP<3$, and $\varpi>35$ mas; green dashed-dotted histogram: sources with $G<12$, $1<BP-RP<2$, and $\varpi>35$ mas; red dashed histogram: sources with $G<12$, $BP-RP<1$, and $\varpi>35$ mas. The median RUWE values for the three distributions are 1.23, 1.06, and 1.04, respectively. Source: \url{https://gea.esac.esa.int/archive/}. 
              }
         \label{fig:ruwedist}
   \end{figure}

In order to resolve this possible ambiguity, one should consider that the RUWE empirical normalisation factor is a function of magnitude and colour\footnote{See \cite{Lindegren2018,Lindegren2021} and the technical note GAIA-C3-TN-LU-LL-124-01 accessible at \url{https://www.cosmos.esa.int/web/gaia/dr2-known-issues\#AstrometryConsiderations}.} that is meant to compensate for a variety of calibration errors (particularly affecting very bright, very blue, and very red sources) in a statistical sense. Its derivation makes use of the full set of single-star solutions, which are and are not well behaved, in any given magnitude-colour bin. Nearby red dwarfs exhibit elevated RUWE values with respect to the equivalent distribution for bluer sources, as shown in Fig. \ref{fig:ruwedist} based on a simple query of the Gaia archive. The underlying stellar samples used for constructing the normalisation factor in the colour bins shown in Fig. \ref{fig:ruwedist} are vastly different, that is dominated by nearby dwarfs (with a mixture of solutions that are and are not well behaved) at the blue end and dominated by distant giants (with a different mixture of solutions) at the red end. This helps to explain the different RUWE distributions shown in Fig. \ref{fig:ruwedist}, with the consequence that the RUWE values for nearby red sources might have to be dealt with while keeping this in mind.  While additional investigations of the issue are beyond the scope of this work, we conclude that the high RUWE for GJ 463 does indicate that the Gaia DR3 time baseline already allows us to see evidence of the presence of GJ 463 b, but the actual departure from a good single-star fit might be less significant than formally indicated, helping to fully reconcile the excess scatter in Gaia DR3 astrometry for the source with the presence of GJ 463 b.

\section{Discussion}

   \begin{figure}
   \centering
$\begin{array}{cc}
\includegraphics[width=0.95\columnwidth]{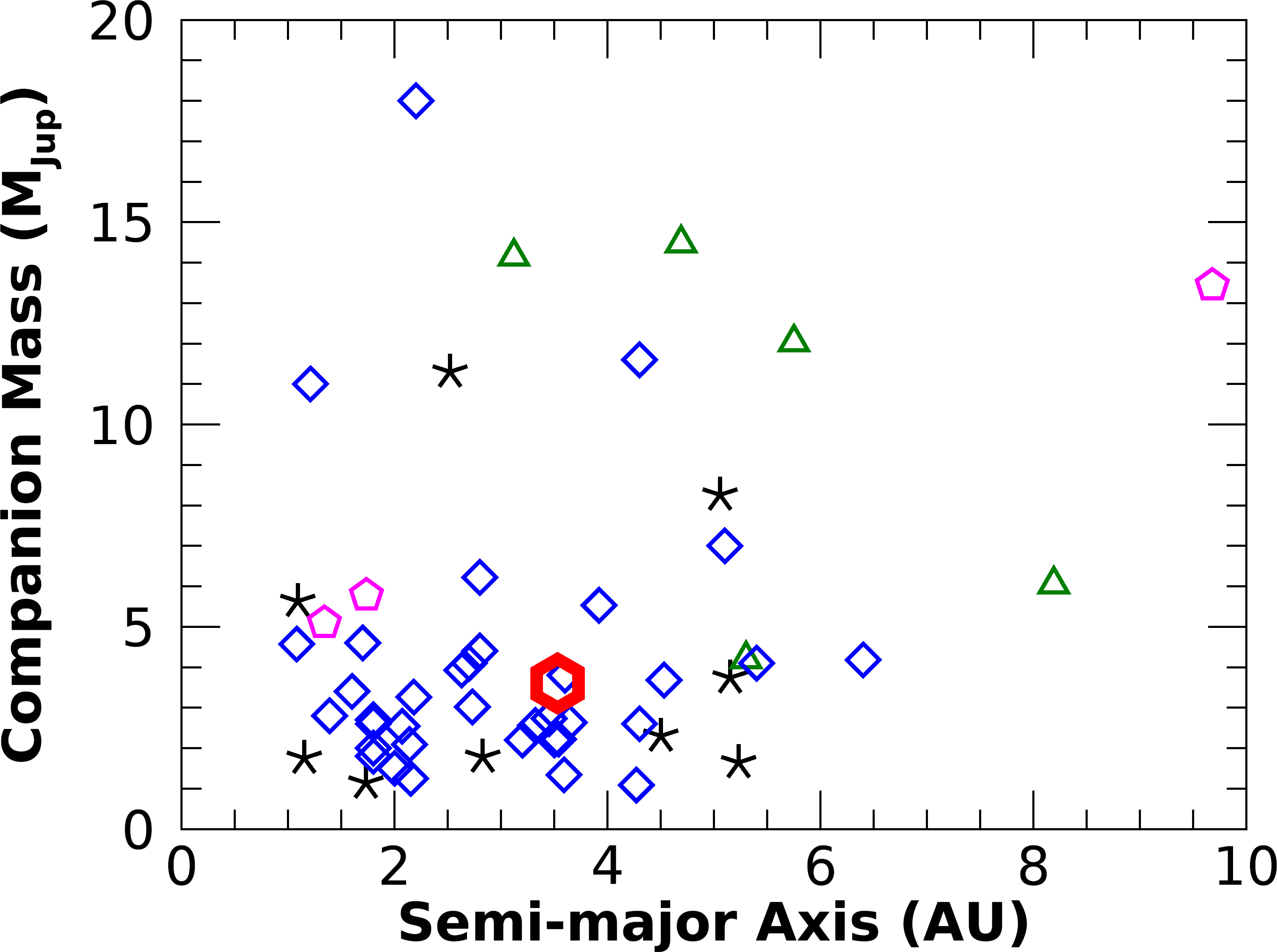} \\ 
\includegraphics[width=0.95\columnwidth]{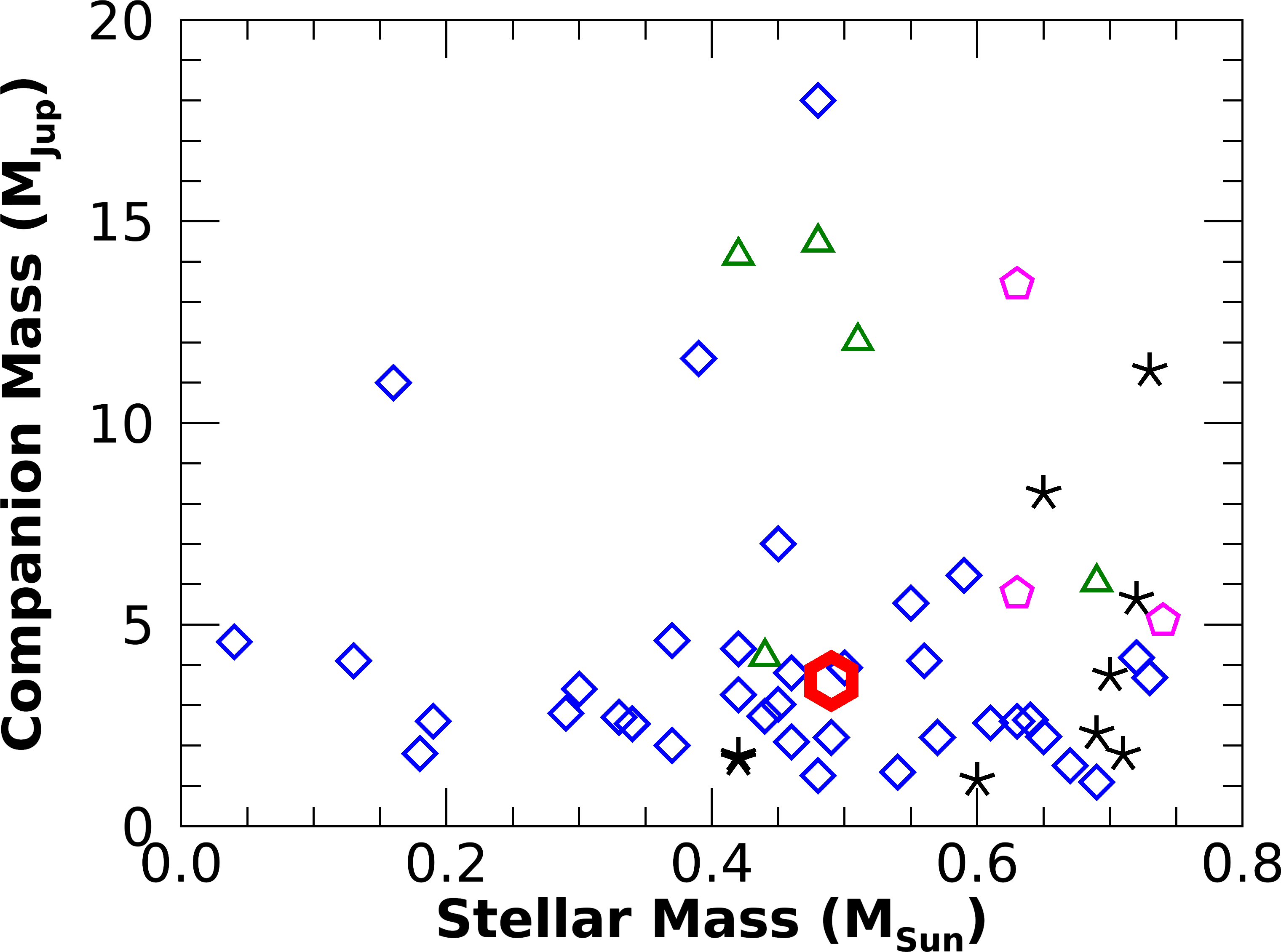} \\
\end{array} $
      \caption{Top: Mass vs orbital separation for companions with mass $1\leq M_c\leq 20$ M$_\mathrm{Jup}$ and $a$ in the range $1-10$ au orbiting stars with $M_\star<0.75$ M$_\odot$. Bottom: Companion mass vs stellar mass. In both panels, black asterisks indicate objects detected by RV techniques, blue diamonds are companions identified in microlensing surveys, green triangles are companions uncovered via eclipse timing variations, and magenta pentagons correspond to objects with true masses determined by joint RV+astrometry analyses. The red hexagon indicates the location of GJ 463 b. Data are from \url{https://exoplanetarchive.ipac.caltech.edu/}.
              }
         \label{fig:comparison}
   \end{figure}

The long-period massive companion GJ 463 b was dubbed by \citet{Endl2022} a 'Jupiter analogue'. The inference from the detailed analysis of absolute astrometry constrained by the published RV solution points towards an object significantly more massive than Jupiter. Recently, super-Jupiters ($1\leq M_p \leq10$ M$_\mathrm{Jup}$) around solar-mass primaries had their true masses constrained via joint RV + astrometric PMA analyses (14 Her b, \citealt{Bardalez2021}; HD 83443 c, \citealt{Errico2022}). GJ 463 b is, to our knowledge, the planetary companion with a dynamical mass constraint based on this methodology orbiting the lowest-mass star known to date. In the regime of intermediate separations $1-10$ au and companions with masses $1-20$ M$_\mathrm{Jup}$ around late-K- and M-dwarf primaries, the true mass value for GJ 463 b appears representative of the majority of objects unveiled by microlensing surveys and eclipse timing variation measurements. 

\begin{figure}
   \centering
   \includegraphics[width=0.95\columnwidth]{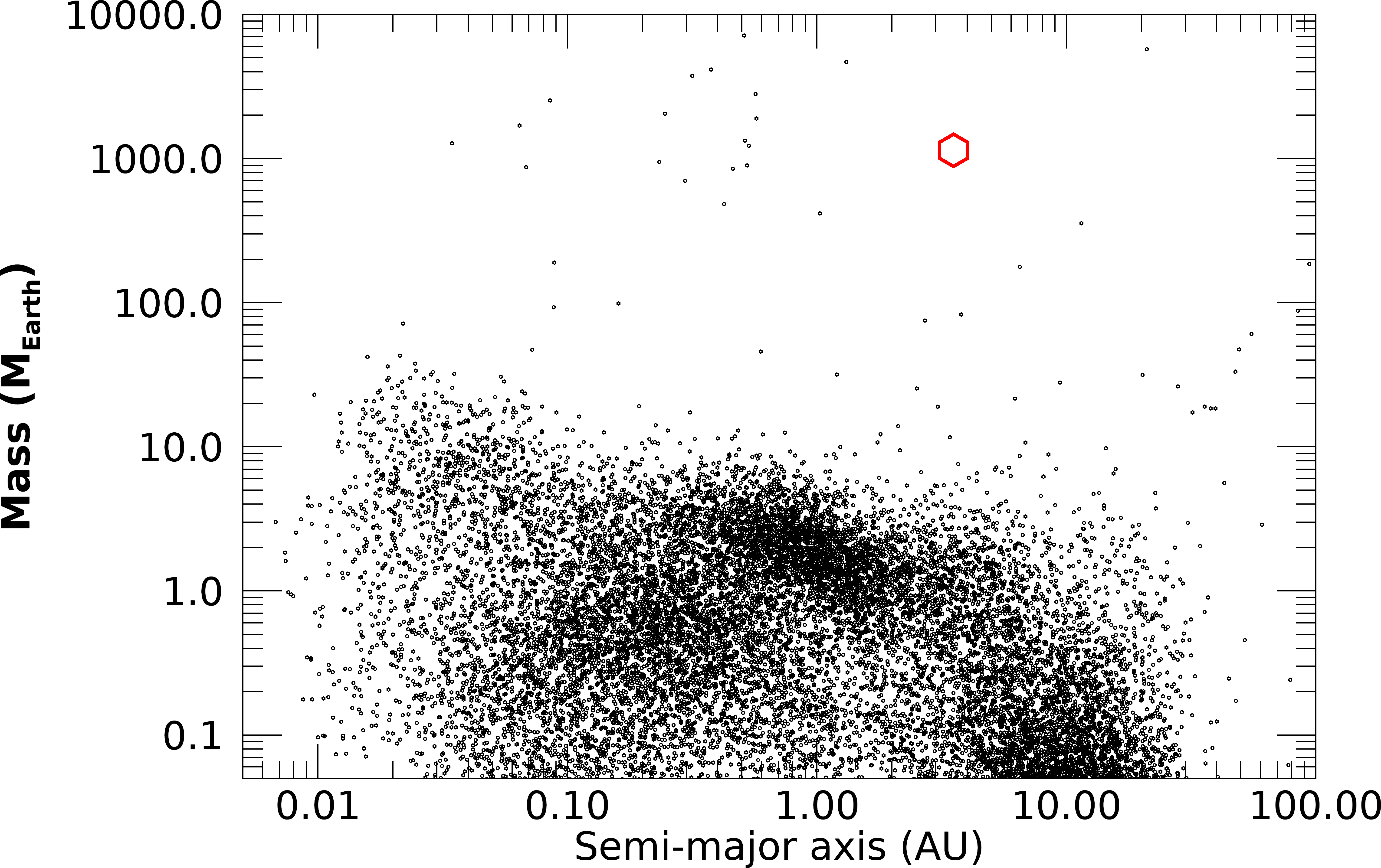}
      \caption{Synthetic population of planets as a function of the orbital separation and mass around a $0.5$-$M_\odot$ primary (simulation SGM11 from \citet{Burn2021}). The large red hexagon identifies the location of GJ 463 b. Source: \url{https://dace.unige.ch/}. 
              }
         \label{fig:bern}
   \end{figure}

High-mass planets are difficult to form around low-mass M dwarfs under the standard paradigm of formation by core accretion \citep{Laughlin2004,Ida2005}. Fig. \ref{fig:bern} shows a synthetic population of planets in the $M_p-a$ plane around a 0.5-$M_\odot$ primary, produced using the Bern global model of planetary formation and evolution \citep{Burn2021}. The population contains only 13 giant planets (0.05\% of the sample) with a mass greater than 1000 M$_\oplus$ ($\sim3.2$ M$_\mathrm{Jup}$), with all but one orbiting at much shorter separations than GJ 463 b. In the \citet{Burn2021} study, no giant planets with masses larger than that of Saturn were produced around $<0.5$-$M_\odot$ primaries. More recently \citet{Schlecker2022} confirmed the difficulty of core-accretion-based population synthesis models in explaining the presence of Doppler-detected giant planets around $\lesssim0.5$-$M_\odot$ stars. In this respect, a system such as that of GJ 463, with a $0.49$-$M_\odot$ primary and a 3.6 M$_\mathrm{Jup}$ companion at 3.5 au, is not easily explained within the context of current core accretion theories, unless additional mechanisms are invoked, such as strong, artificial inhibition of orbital migration or very fast core growth via the accretion of sufficiently high centimetre-sized pebble fluxes (see \citet{Schlecker2022}, and references therein). Provided a sufficiently large disk-to-star mass ratio ($\gtrsim0.3$), such objects are ideal candidates for formation by disk instability: The conditions for the instability to happen are met in the outer disk regions, well beyond the snow line (e.g. \citealt{Boss2006,Stamatellos2007,Backus2016,Mercer2020}). Scenarios of subsequent dynamical evolution, such as migration towards the star due to disk-planet or planet-planet interactions (see, e.g. \citet{Mercer2020}, and references therein), could help explain the presence of the population of companions similar to GJ 463 b at intermediate separations from $\lesssim0.5$-$M_\odot$ primaries. Knowledge of their exact orbital properties and true masses is vital in order to carry out the most robust studies of their demographics. In this respect, GJ 463 b significantly adds to the small lot of objects (highlighted in Fig. \ref{fig:comparison}) with joint constraints on their true masses from RVs and astrometry, that is GJ 676 A b,c \citep{Sahlmann2016,Feng2022}, and BD-17 0063 \citep{Arenou2022,Winn2022}. In this sample, GJ 463 b orbits the lowest-mass primary. 

Not surprisingly, the vast majority of the candidate substellar companions whose astrometric orbits were recently published with Gaia DR3 have been identified around nearby M dwarfs \citep{Arenou2022,Holl2022}. The orbital periods of this sample are almost invariably $\lesssim1000$ d, which is a consequence of the DR3 data time span. For the sub-sample of candidates with masses $\lesssim20$ M$_\mathrm{Jup}$, the median distance, primary mass, and orbital periods are $\sim20$ pc, $\sim0.39$ M$_\odot$, and $\sim540$ d, respectively. While a detailed Gaia DR3 survey sensitivity analysis for such companions has yet to be performed, we have seen (Fig. \ref{fig:sensitivity}) that a system with the characteristics of GJ 463 should have been marginally detectable. Indeed, GJ 463 passed the RUWE$=1.4$ threshold formally adopted by \citet{Halbwachs2022} in the Gaia DR3 astrometric binary star processing. The fact that no non-single star solution for GJ 463 is present in the Gaia DR3 archive indicates that any attempted solution must not have passed the rather stringent acceptance thresholds, which were imposed in order to cope with significant fractions of spurious solutions, at the expense of possibly failing to recognise bona fide perturbations induced by real companions (see \citealt{Arenou2022,Halbwachs2022} for details). The expectation is that Gaia astrometry alone will be able to deliver a complete orbital solution and significantly improve the true mass estimate for GJ 463 b with the publication of Gaia DR4, slated to be released by the end of 2025 and based on 66 months of data collection.

\begin{acknowledgements}

We gratefully acknowledge the referee, T. Brandt, for a swift report with constructive criticisms and stimulating suggestions that quantitatively improved an earlier version of this manuscript. 
A.S. is grateful to F. Arenou for useful discussions. This work has made use of data from the European Space Agency (ESA) mission {\it Gaia} (\url{https://www.cosmos.esa.int/gaia}), processed by the {\it Gaia} Data Processing and Analysis Consortium (DPAC,
\url{https://www.cosmos.esa.int/web/gaia/dpac/consortium}). Funding for the DPAC has been provided by national institutions, in particular the institutions participating in the {\it Gaia} Multilateral Agreement. A.S. acknowledges the financial support from the agreement ASI-INAF n.2018-16-HH.0, and from the Italian Space Agency (ASI) under contract 2018-24-HH.0 "The Italian participation to the Gaia Data Processing and Analysis Consortium (DPAC)" in collaboration with the Italian National Institute of Astrophysics. This work has made use of the SIMBAD and VizieR databases and catalog access tool at the CDS, Strasbourg (France), and NASA’s Astrophysics Data System Bibliographic Services.
This publication makes use of the Data \& Analysis Center for Exoplanets (DACE), which is a facility based at the University of Geneva (CH) dedicated to extrasolar planets data visualisation, exchange and analysis. DACE is a platform of the Swiss National Centre of Competence in Research (NCCR) PlanetS, federating the Swiss expertise in Exoplanet research. The DACE platform is available at \url{https://dace.unige.ch}. This research has made use of the NASA Exoplanet Archive, which is operated by the California Institute of Technology, under contract with the National Aeronautics and Space Administration under the Exoplanet Exploration Program.
\end{acknowledgements}

%
   \bibliographystyle{aa} 
   \bibliography{GJ463b_biblio} 
%

\end{document}